# Bohr and von Neumann on the Universality of Quantum Mechanics: Materials for the History of the Quantum Measurement Process

Federico Laudisa

Department of Humanities and Philosophy, University of Trento

### Abstract

The Bohr and von Neumann views on the measurement process in quantum mechanics have been interpreted for a long time in somewhat controversial terms, often leading to misconceptions. On the basis of some textual analysis, I would like to show that – contrary to a widespread opinion – their views should be taken less inconsistent, and much closer to each other, than usually thought. As a consequence, I claim that Bohr and von Neumann are conceptually on the same side on the issue of the universality of quantum mechanics: hopefully, this might contribute to a more accurate history of the measurement problem in quantum mechanics.

## 1    Introduction

The role played by the notion of measurement in the development of the Bohrian interpretation of quantum mechanics hardly needs to be emphasized. Over and above the subtleties in the exegesis of the Bohr writings, in the Bohrian conceptual framework it is the measurement context that appears to provide a meaning for physical experience and, within this experience, for the meaningful attribution of properties to quantum systems. Still, the Bohr writings themselves do *not* contain any explicit and detailed – albeit theoretical – *model* of the measurement process. The opening lines of his 1928 complementarity article express effectively his overall attitude to address foundational problems in non-formal ways[1], an attitude on which a leading interpreter has aptly commented: "A thorough formal theory of measurement, either in classical or quantum physics was, to his mind, out of the question, since any axiomatization would depend on primitive descriptive concepts drawn from the language justified by ordinary experience." (Honner , p. 65).

---

[1] "I shall try, *by making use only of simple considerations and without going into any details of technical mathematical character*, to describe to you a general point of view which I believe is suited to give an impression of the general trend of the interpretation of the theory from its very beginning" (Bohr 1934, p. 52).

In any case, it was rather clear from the 1928 paper onwards that the Bohrian interpretation of quantum mechanics would have reserved for the measurement process a special role. This circumstance would have soon induced physicists to investigate the relation between the classical realm and the quantum one: in the early days of quantum mechanics, a measurement started to be conceived as a very special sort of interaction between pairs of systems, one member of which is described by classical physics while the other is described by quantum physics. How is this classical-quantum relation supposed to be accounted for? The first textbooks on quantum mechanics of the twentieth century appeared few years after the Bohr complementarity paper: *Die Physikalischen Prinzipien der Quantentheorie* by Werner Heisenberg and *The Principles of Quantum Mechanics* by P.A.M. Dirac, both published in 1930. Both texts aspired to provide a general framework, as rigorous as possible, for quantum mechanics but neither included an explicit analysis of the measurement process in terms of the *universality* of quantum mechanics: in light of the 'quantumness' of the natural world at the microscopic scale, why should quantum mechanics not govern *all* natural processes, including macroscopic instruments involved in a typical quantum measurement? If it does, what are the presuppositions for the theory to account for the measurement process and what the consequences for the interpretation of the theory itself?

In retrospective, it is the Bohrian doctrine concerning this relation that in fact appears to be the first to address this issue. As a matter of fact, however, the issue of whether or not *according to Bohr* quantum mechanics should be taken to be universal has been a matter of dispute in the Bohrian scholarship. Bohr repeatedly argues in his works that *we cannot do without* classical concepts in accounting for the experimental evidence in quantum measurements, but does this acknowledgment imply that the Bohrian overall foundational outlook legitimates a *non*-universalistic interpretation of quantum mechanics? According to an old-fashioned reading, it does. Moreover, in this reading the classical/quantum 'cut' in the description of measured systems *vis-à-vis* measuring instruments appears to suggest a truly *ontological* 'cut' between a classical world and a quantum world.

This reading has been rather popular for a long time, especially in the community of physics proper rather than in the community of history and philosophy of physics but, to be fair, the historico-philosophical scholarship on Bohr had been questioning more and more this reading, in favour of a classical/quantum cut of a purely pragmatic or epistemological character. The alleged unavoidability of classical concepts, in order to describe what goes on ordinarily at the end of an experiment in a laboratory, need not imply that quantum mechanics is not universal, and is rather to be justified by the constraints of *communication* and *inter-subjectivity*. According to Dugald Murdoch, for instance "Bohr's point is that it is a necessary condition of *unambiguous* and *objective* communication that the experimental apparatus be describable in common-sense ordinary-language terms" (Murdoch 1987, p. 100) and, in more recent times,

Dennis Dieks argued that "[...] the idea that Bohr denied the universal validity of quantum mechanics is mistaken. Indeed, from his earliest writings on Bohr argues from the assumption that quantum theory *is* universal, in the sense of applicable to both micro and macro systems [...] the quantum description remains ontologically primary, even when we use (as we must) classical concepts to describe macroscopic objects." (Dieks 2017, pp. 312-313).

The aim of the present paper, however, is not to adjudicate the problem whether the Bohrian interpretation of the classical/ quantum relation should be read in pragmatic or ontological terms. Our question is different, and more historically-oriented. The quantum physics community had to wait for *Die Mathematische Grundlagen der Quantenmechanik* by Jón (John) von Neumann in 1932 to see an entire chapter *explicitly* devoted to the development of a formal model of a quantum measurement. In the von Neumann treatment (although, as we will see, in a non completely transparent way), he explicitly confronts the implications of the assumption that – in the context of a measurement of a physical quantity on a quantum system *S* with an apparatus *A* – the laws of QM govern *both S and A*, an assumption that, without additional requirements, entails the emergence of the measurement problem: now, did ever Bohr discuss explicitly the issue of the universality of quantum mechanics on the background of the von Neumann formal context? More generally what is the connection, if any, between the Bohr view of quantum measurement and the framework for the quantum measurement process that von Neumann elaborated in his 1932 treatise? It is this point that will be the target of the present paper; the analysis of this point – an analysis that, to my knowledge, is still lacking – might contribute to the project of a more documented history of the measurement process in the development of quantum mechanics. Moreover, the connection Bohr-von Neumann turns out to be interesting also because von Neumann himself has been long credited with a view whose textual support is controversial, namely the view according to which the infamous 'collapse' of the wave function is a genuinely physical process. In particular, I conjecture that the Bohr claim on the role of classical concepts as a *pragmatic* recipe, necessary to account for the emergence of a definite outcome at the end of a measurement process, in fact resonates quite closely with a literal understanding of the relevant sections of the von Neumann treatise on the measurement process; in this vein, the views of both Bohr and von Neumann turn out to have suffered from specular misunderstandings and misconceptions and, according to the analysis carried out in the present paper, they appear to be much closer than many have thought in the past. According to my interpretation, the Bohr view of the quantum measurement would turn out to be perfectly consistent with the universality of quantum mechanics, taken for granted by von Neumann, showing *also* on the Bohr side a growing awareness that the times were ripe for addressing the idea that the measurement process – as a *natural process in itself* – was to be conceived as an entirely *quantum* process.

The plan of the paper is the following. In section 2 I will review the ways in which Heisenberg in *Die Physikalischen Prinzipien der Quantentheorie* and Dirac in *The Principles of Quantum Mechanics* deal with the measurement process in quantum mechanics in their textbooks. This section is focused on these works since they can be taken to represent what was considered at the time the state of the art, as far a systematic formulation of quantum theory was concerned. As will be argued, the theoretical landscape provided by these two works does not seem to show a clear awareness of the implications that may follow for the foundations of the theory from treating measuring instruments as quantum systems in their own right. The reasons for this lack of awareness are different in the two books: the role of a Copenhagenish view acting on the background in the case of Heisenberg, the pragmatic attitude toward foundational issues in the case of Dirac. In the section 3, I will shortly recall the main elements of the Bohr analysis of the quantum measurement process, with an emphasis on two particular points. First, Bohr appears to be the first to specifically and thematically address the relation classical/quantum in the quantum measurement process as a foundational issue in its own right. Second, the Bohrian conceptual framework can be interpreted so as to support a *coexistence* of the use of the language of classical physics on one side – in order to account for the outcomes of the quantum measurement process – and the universality of quantum mechanics as a theory governing the whole physical world on the other. This claim of coexistence paves the way to a comparison of the views of Bohr and von Neumann – in the section 4 – in spite of the surprisingly meager evidence concerning the *actual* interactions between the two. After recalling the von Neumann model of the measurement process (sketched in *Die Mathematische Grundlagen der Quantenmechanik*), for the comparison I will rely essentially on a single paper – "The causality problem in atomic physics" – that Bohr presented at a conference held in 1938 in Warsaw: this conference was attended also by von Neumann and appears to be the only public exchange between them. As I will attempt to argue, the Bohr 1938 paper appears to show that Bohr himself is aware of the possibility to rely on the von Neumann model of the measurement process in order to justify more robustly the above mentioned coexistence of the use of a classical language *vis-à-vis* the universality of quantum mechanics: this circumstance makes the Bohr analysis of the quantum measurement in the Thirties much less alien to the direction that the history of the quantum measurement process would have taken later than previously thought. I will finally draw some conclusions in the final section.

## 2    The Measurement Process in the Textbooks of Werner Heisenberg and Paul A. M. Dirac

In the present section I will shortly review the attitude toward the measurement process in quantum mechanics that surfaces in two very influent textbooks, appeared in the heroic times of the birth and consolidation of quantum mechanics: Werner Heisenberg's *Die Physikalischen Prinzipien der Quantentheorie* and Paul A.M. Dirac's *Principles of Quantum Mechanics*, both published in 1930. Since the authors are among the giants of the twentieth century theoretical physics, a warning is in order. The aim of this section is limited to hint to a pair of highly qualified instances, where the quantum measurement process is interpreted on the background of a classical/quantum interaction in which the measuring apparatus side is still taken to belong unquestionably and intrinsically to the *classical* realm. Given this state of affairs, the 'quantumness' of the quantum measurement process is confined to the circumstance according to which the measuring apparatus 'disturbs' the quantum system, waiting to be measured, to an extent which is unconceivable in classical terms. The analysis I propose in the present section is meant to emphasize two points, both functional to the Bohr and von Neumann contributions, considered in the next sections. First, both texts employ (although at a different degree) a 'disturbance' view of the quantum measurement, a view that might have been considered close in spirit to some passing remarks in the Bohr 1928 complementarity paper but that, in fact, is inconsistent both with other main principles of quantum mechanics and with the mature views of Bohr, concerning in particular the context-dependence of properties for quantum systems in general. Second, the endorsement of this view by Heisenberg and Dirac clearly shows that their interpretation of the measurement process leave the issue of the universality of quantum mechanics totally unqualified, namely, that at that stage neither of them takes the issue of describing quantum-mechanically *both* the measured system and the measuring apparatus to be a genuine issue, that is, a problem that *needs* to be addressed in order to assess the scope of the evolving quantum theory. Only in the Bohr analysis of the classical/quantum interaction, no matter how controversial and involved it might appear, the issue starts officially to occupy the position it deserves in the debate on the foundations of quantum mechanics.

Heisenberg's *Die Physikalischen Prinzipien der Quantentheorie* derived from a set of lectures that Heisenberg delivered at the University of Chicago in 1929. Already from the Preface Heisenberg emphasizes his intellectual debt to Niels Bohr:

> The lectures that I gave at the University of Chicago in the spring of 1929 afforded me the opportunity of reviewing the fundamental principles of quantum theory. Since *the conclusive studies of Bohr in 1927* there have been no essential changes in these principles, and many new experiments have confirmed important consequences of the theory [...] On the whole the book contains nothing that is not to be found in previous publications, *particularly in the investigations of Bohr*. The purpose of the book seems to me to be fulfilled if it contributes somewhat to the diffusion of that "Kopenhagener Geist der Quantentheorie", if I may so express myself, which has directed the entire development of modern atomic physics. (Heisenberg 1930, Preface, emphasis added).[2]

In the introductory chapter Heisenberg refers to several issues that have deep implications by the foundational point of view and that testify the above mentioned Bohrian inspiration of the whole enterprise. First, the emphasis on common language: a tool which is indispensable on one side, in order to describe and effectively handle the experimental results of physical procedures but whose expressive limitations, on the other side, may lead us to insurmountable problems.

> The experiments of physics and their result can be described in the language of daily life. Thus if the physicist did not demand a theory to explain his result and could be content, say, with a description of the lines appearing on photographic plates, everything would be simple and there would be no need of an epistemological discussion. Difficulties arise only in the attempt to classify and synthesize the results, to establish the relation of cause and effect between them – in short, to construct a theory. (Heisenberg 1930, p. 1)

Second, a new formulation of what are the general conditions that a satisfactory theoretical scheme for quantum phenomena should obey. In the opening sentence of the celebrated 1927 paper that contains the first formulation of the uncertainty relations, Heisenberg had written:

> We believe to have understood a physical theory *intuitively* if we can imagine the experimental consequences of the theory qualitatively in all simple cases, and if, at the same time, we have recognized that the application of the theory will never contain internal contradictions." (Heisenberg 1927, p. 172)[3]

This characterization of what we require from a physical theory when we are confronted with the new quantum phenomena is somehow recalled in the 1930 book, but with a new, interesting twist. At the end of the section 1 of the Introduction Heisenberg writes:

---

[2] In the 1938 paper that we will discuss in detail in the section 4, Niels Bohr explicitly writes that in this book "typical examples of measuring processes [...] have been treated in detail" (Bohr 1938, p. 100).

[3] I use here the translation included in a paper by R.F. Werner and T. Farrelly, devoted to a thorough analysis of the 1927 Heisenberg paper (Werner, Farrelly 2019, p. 463); the widespread English translation of the Heisenberg paper, due to J.A. Wheeler and W.H. Zurek and contained in their 1983 collection of classic papers on the foundations of quantum mechanics (Wheeler, Zurek 1983), mistakenly cancels both from the title and the opening sentence the German adjective *anschaulich* ('intuitive') which, on the contrary, is highly loaded from a conceptual point of view. On the connotation of *anschaulich* in the opening sentence of the Heisenberg paper, Werner and Farrelly aptly comment: "This quotation, which Heisenberg also used in his later years, is remarkably modern. The term *Anschauung* (literally 'looking at something') is stripped here of almost all connotations of imagining a scene or a picture. Like the 'internal virtual images' of Hertz and Galilei's geometrical figures as letters in the book of nature, it can just as well refer to an intuition about an algebraic or logical structure. Whether this widening of the concept of *Anschaulichkeit* convinced Heisenberg's contemporary critics, however, is questionable." (Werner, Farrelly 2019, pp. 463-464). On the Heisenberg conceptual shift concerning the notion of *Anschaulichkeit*, see also the section 3.4 of Camilleri 2009, entitled "Redefining *Anschaulichkeit*" (and the related bibliography therein).

The program of the following considerations will therefore be: first, to obtain a general survey of all concepts whose introduction is suggested by the atomic experiments; second, *to limit the range of application of these concepts; and third, to show that the concepts thus limited, together with the mathematical formulations of quantum theory, form a self-consistent scheme.* (Heisenberg 1930, p. 4, emphasis added).

According to the most qualified scholarship, the italicized passage testifies of a change of perspective with respect to the 1927 paper, induced by the discussions that Heisenberg had with Bohr between the 1927 paper and the 1930 book. Kristian Camilleri, for instance, argues convincingly that the 1927 Heisenberg paper assumes a sort of verificationist criterion of meaning, which – given the fixation of a value for the position [momentum] of the particle at a given time $t$ – would make the notion of momentum [position] of that particle at $t$ simply *meaningless* (Camilleri 2009, p. 94) [4]. Still according to Camilleri, Heisenberg changes his mind as a consequence of the reading of the Bohr 1927 paper and the ensuing exchange of ideas with Bohr himself, so that, starting already from the 1930 book, Heisenberg agrees on the *indispensability* of classical concepts in quantum mechanics, which is a major claim of Bohr. So, according to the italicized passage, it is not that some classical concepts – the most notable of which are position and momentum – are just deprived of meaning by the new quantum theory; rather, the new quantum theory shows that the range of application of our familiar notions of some classical concepts, notions whose meaning we continue to have a good grasp of, is irreducibly restricted.[5]

Moreover, the Heisenberg 1930 book suggests an oscillation between two views that appear in a mutual tension, if not inconsistent: one of these views, as we shall see, emerges clearly in the formulation chosen by Dirac in his 1930 book. In fact Heisenberg, on one side, appears to suggest the endorsement of what has been called a 'disturbance view' of measurement, "according to which it is only the perturbation brought about by the act of observation that precludes us from knowing the precise position and momentum [of the quantum particle at hand]" (Camilleri 2009, p. 105). In a thorough analysis proposed already in 1981, Harvey Brown and Michael Redhead had described this view as follows:

In the case of the microscope experiment, [...] an incoming light particle (which may be considered the measuring system) collides with an electron (the object system), the collision being governed by the classical laws of conservation of energy and linear momentum. In both cases, the light particle is endowed with quantal fluctuations which correspond to the use of the Einstein-de Broglie relations for the energy and momentum for these systems. *The electron may be effectively considered a classical particle, which in interaction with the quantal measuring agent gains* *quantal fluctuations, as witnessed by the final indeterminacy relations obtained for it.* (Brown, Redhead 1981, p. 2)

This view appears to be in fact at work in the Heisenberg 1930 book. In the very first pages of the Introduction, distinguishing relativity and quantum theory in an important respect, Heisenberg writes:

> Although the theory of relativity makes the greatest of demands on the ability for abstract thought, still it fulfills the traditional requirements of science in so far as it permits a division of the world in subject and object (observer and observed) and hence a clear formulation of the law of causality. [...] Particularly characteristic of the discussions to follow is the interaction between observer and object; *in classical physical theories it has always been assumed either that this interaction is negligibly small, or else that its effect can be eliminated fro the result by calculations based on "control" experiments. This assumption is not permissible in atomic physics; the interaction between observer and object causes incontrollable and large changes in the system being observed, because of the dicontinuous changes characteristic of atomic processes.* (Heisenberg 1930, pp. 2-3)[6]

In a subsequent section (entitled *Illustrations of the uncertainty relations*) Heisenberg again characterizes the constraints induced by the relations in terms of limitation to the *knowledge* of properties which, in principle, are possessed by the quantum particle:

> The uncertainty principle refers to the degree of indeterminateness in the possible present knowledge of the simultaneous values of various quantities with which the quantum theory deals; it does not restrict, for example, the exactness of a position measurement or a velocity measurement alone. Thus suppose that the velocity of a free electron is precisely known, while the position is completely unknown. Then the principle states that every subsequent observation of the position will alter the momentum by an unknown and undeterminable amount such that after carrying out the experiment our knowledge of the electronic motion is restricted by the uncertainty relation. (Heisenberg 1930, p. 20).

Furthermore, in a section devoted the Bohrian notion of complementarity, Heisenberg indicates the possibility of observing a physical event without disturbing it appreciably as the mark of classical physics (relativity included), suggesting that this is exactly the most fundamental novelty introduced by quantum theory:

> [...] the resolution of the paradoxes of atomic physics can be accomplished only by further renunciation of old and cherished ideas. Most important of these is the idea that natural phenomena obey exact laws – the principle of causality. In fact, our ordinary description of nature, and the idea of exact laws, rests on the assumption that it is possible to observe the phenomena without appreciably influencing them. To co-ordinate a definite cause to a definite effect has sense only when both can be observed without introducing a foreign element disturbing their interrelation. The law of causality, because of its very nature, can only be defined for isolated systems, and in atomic physics even approximately isolated systems cannot be observed. This might have been foreseen, for in atomic physics we are dealing with entities that are (so far as we know) ultimate and indivisible. (Heisenberg 1930, pp. 62-63).

---

[6] The relativity/quantum theory relation is another point on which the 1927 uncertainty paper and the 1930 book diverge: in the former Heisenberg shapes the discussion on his interpretation of the 1905 Einstein paper, suggesting essentially that his analysis on the uncertainty relation does for quantum theory what the Einstein analysis of simultaneity did for relativity, whereas in the latter the two theories are made to diverge essentially on the observer-observed relation: in this respect, relativity would be 'classical' and quantum theory would be 'non-classical' (on the point see again Camilleri 2009, pp. 94-95).

On the other hand the disturbance view, which was *not at all* shared by Bohr [7], appears to coexist for Heisenberg with a different view, expressed in certain places, according to which the very idea that quantum particles *do* have a definite position and momentum before measurement is to be rejected. In the chapter 2 of the 1930 book Heisenberg refers to the indeterminateness in the value of velocity of an electron, as a consequence of a precise determination of the value for its position, as "an *essential* characteristic of the electron" (Heisenberg 1930, p. 14, emphasis added), whereas in a paper published in 1931 on *Monatshefte für Mathematik und Physik* (but based on a lecture given in December 1930, Camilleri 2009, p. 106) Heisenberg argues that "the uncertainty relations hence should not simply be conceived of as the impossibility of of precisely knowing or measuring the position and velocity [of an electron]; the uncertainty relations signify that an application of the words "position, velocity" loses any reasonable meaning beyond specified limits." (Heisenberg 1931, p. 367, quoted from Camilleri 2009, p. 106).

The role of a disturbance view of measurement is also apparent in the 1930 Dirac book *The Principles of Quantum Mechanics*. In the section 1 (*The need for a quantum theory*) of the chapter 1, Dirac shortly describes the wave/particle duality and, emphasizing that this dual nature is not confined to light but is typical of any material particle, recalls that "we have here a very striking and general example of the breakdown of classical mechanics – not merely an inaccuracy in its laws of motion, but *an inadequacy of its concepts to supply us with a description of atomic events*." (Dirac 1958[4], p. 3, emphasis in the original). According to Dirac, the need of a new theory in order "to account for the ultimate structure of matter" is based not only on experimental evidence but also on "general philosophical grounds", namely the necessity to provide (what Dirac takes to be) an *absolute* criterion to distinguish *big* and *small*:

So long as *big* and *small* are merely relative concepts, it is no help to explain the big in terms of the small. It is therefore necessary to modify classical ideas in such a way as to give an absolute meaning to size." (Dirac 1958[4], p. 3).

---

[7] "Indeed Bohr himself in his later writings retreated from the disturbance doctrine as an explanation of the uncertainty relations, emphasizing the wholeness of a quantal phenomenon involving the specification of the experimental arrangement in classical terms, rather than mechanical transmission of uncontrollable disturbance as the source of the characteristic features of the theory. This shift was apparently due to difficulties of understanding the Einstein-Polodsky-Rosen discussion in terms of a disturbance theory." (Brown, Redhead 1981, p. 3)

In the case of Dirac, it is his search for an absolute criterion for the determination of size that motivates the adoption of a disturbance view of observation and measurement in quantum mechanics. This is how such view plays the desired role in the Dirac presentation:

> In order to give an absolute meaning to size, such as is required for any theory of the ultimate theory of matter, we have to assume that *there is a limit to the fineness of our powers of observation and the smallness of the accompanying disturbance – a limit which is inherent in the nature of things and can never be surpassed by improved technique or increased skill on the part of the observer*. If the object under consideration is such that the unavoidable limiting disturbance is negligible, then the object is big in the absolute sense and we may apply classical mechanics to it. If, on the other hand, the limiting disturbance is not negligible, then the object is small in the absolute sense and we require a new theory for dealing with it. (Dirac 1958[4], pp. 3-4, emphasis in the original).[8]

As we stressed above in the case of the Heisenberg 1930 book, this disturbance view is clearly in tension with the claim, *also* based on the principles of quantum mechanics, according to which quantum particles *do not* have definite properties before measurement. In the Dirac 1930 book this tension is apparent, for instance, in the description of the particular features that quantum theory displays when applied to the phenomenon of the superposition of polarization states for photons, presented in the section 2 (*The polarization of photons*) of the same chapter 1.

Dirac describes a now familiar situation: a beam of light is directed toward an apparatus (a tourmaline crystal), that is designed to transmit entirely light polarized perpendicularly to a given axis and to absorb entirely light polarized parallel to that axis. The notorious problem is how to describe what happens when the polarization of the incident beam is neither perpendicular nor parallel to the given axis:

> A beam that is plane-polarized in a certain direction is to be pictured as made of photons each plane-polarized in that direction. This picture leads to no difficulty in the cases when our incident beam is polarized perpendicular or parallel to the optic axis. We merely have to suppose that each photon polarized perpendicular to the axis passes unhindered and unchanged through the crystal, and each photon polarized parallel to the axis is stopped and absorbed. A difficulty arises, however, in the case of the obliquely polarized indicent beam. Each of the incident photons is then obliquely polarized and it is not clear what will happen to such a photon when it reaches the tourmaline. (Dirac 1958[4], p. 5)

---

[8] This passage is followed in the Dirac book by a remark on the implication of the preceding discussion on the idea of causality, which is essentially a reproduction (nearly word-by-word) of the remark on causality by Heisenberg in the 1930 book we quoted above, taken from a section devoted to the Bohrian notion of complementarity.

Dirac takes it necessary to specify what it means to raise a question in terms of 'what will happen' to the single photon, and this allows him to express explicitly an article of operational faith:

> A question about what will happen to a particular photon under certain conditions is not really very precise. To make it precise one must imagine some experiment performed having a bearing on the question and inquire what will be the result of the experiment. *Only questions about the results of experiments have a real significance and it is only such questions that theoretical physics has to consider.* (Dirac 1958[4], p. 5).

In the situation at hand, we imagine to send a single photon at a time toward the crystal and observe whether it passes or not. Since the experiment assumes a 'real significance' only in statistical terms, the experiment will be repeated many times:

> According to quantum mechanics the result of the experiment will be that sometimes one will find a whole photon [...] on the back side and other times one will find nothing. [...] If one repeats the experiment a large number of times, one will find the photon on the back side in a fraction $sin^2\ \alpha$ [where a is the angle to the axis at which the photon is polarized] of the total number of times. Thus we may say that the photon has a probability $sin^2\ \alpha$ of passing through the tourmaline and appearing on the back side polarized perpendicular to the axis and a probability $cos^2\ \alpha$ of being absorbed. *These values for the probabilities lead to the correct classical results for an incident beam containing a large number of photons.* (Dirac 1958[4], pp. 5-6)

Now it is quite clear that here Dirac is adopting a completely operational attitude, putting into brackets the question of whether the photon *does* possess or not a definite polarization property *before* the experiment. This attitude is reinforced by the methodological remark that follows the above statement. Dirac recalls that the result of an experiment "is not determined, as it would be according to classical ideas, by the conditions under the control of the experimenter", but he emphasizes:

> Questions about what decides whether the photon is to go through or not and how it changes its direction of polarization when it does go through *cannot be investigated by experiment and should be regarded as outside the domain of science*." (Dirac 1958[4], p. 6, emphasis added).

This attitude should then lead us to think that – according to Dirac – what quantum mechanics suggests us to do is just collecting the outcomes of experiments and making up the relevant statistics on that basis, refraining from ascribing any definite property whatsoever to photons in states that are neither perpendicular nor parallel to the given axis. Surprisingly, few lines later, Dirac says something that hardly coheres with the above. Introducing the crucial notion of superposition of states, Dirac present it not as a tool to succintly describe the state of photons that give rise to the above mentioned statistics, but as a situation in which the *single* photon *does* have a property – albeit weird and highly 'non-classical' – that of being *simultaneously* in a perpendicular *and* parallel polarization state, although each only to a *partial* extent:

> The further description provided by quantum mechanics runs as follows. It is supposed that a photon polarized obliquely to the optic axis may be regarded as being partly in the state of polarization parallel to that axis and partly in the state of polarization perpendicular to that axis. (Dirac 1958[4], p. 6, emphasis added).

Given this attribution, it is then natural to describe the outcome of measurement via a kind of 'realistic' collapse process:

> When we make the photon meet a tourmaline crystal, we are subjecting it to an observation. We are observing whether it is polarized parallel or perpendicular to the optic axis. The effect of making this observation is to force the photon entirely into the state of parallel or entirely into the state of perpendicular polarization. *It has to make a sudden jump from being partly in each of these two states to being entirely in one or other of them*. (Dirac 1958[4], p. 7, emphasis added).

Clearly, should Dirac not interpret the pre-measurement photon as *having* a property, though weird and not amenable to any classical intuition, he could not talk meaningfully of a 'jump' determined by the collapse, whose action is that of turning the weird <partly perpendicular + partly parallel>-property into one of the two familiar <perpendicular>-property and <parallel>-property. Interestingly, with a further oscillation, some pages later Dirac seems to be willing to deflate what, according to the above passage, appears as a peculiar physical process. In the section 4 of the chapter 1, aptly entitled *Superposition and indeterminacy*, Dirac addresses a possible sceptical reaction to the weirdness of the above process. According to this reaction,

> a very strange idea has been introduced – the possibility of a photon being partly in each of two states of polarization, or partly in each of two separate beams – but even with the help of this strange idea no satisfying *picture* of the fundamental single-photon has been given. (Dirac 1958[4], p. 10, emphasis added).

The Dirac reply is *tranchant*, and entirely on the operational side:

> In answer to the first criticism it may be remarked that the main object of physical science is not the provision of pictures, but is the formulation of laws governing the phenomena and the application of these laws to the discovery of new phenomena. If a picture exists, so much the better; but whether a picture exists or not is a matter of only secondary importance. In the case of atomic phenomena no picture can be expected to exist in the usual sense of the word 'picture', by which is meant a model functioning essentially on classical lines. (Dirac 1958[4], p. 10, emphasis added).[9]

---

[9] Following this very passage, Dirac makes this remark: "One may, however, extend the meaning of the word 'picture' to include any *way of looking at the fundamental laws which makes their self-consistency obvious*. With thi extension, one may gradually acquire a picture of atomic phenomena by becoming familiar with the laws of quantum mechanics." (Dirac 1958[4], p. 10, emphasis in the original). As we can see, it is a remark extremely close to the spirit

# 3   Bohr and the Role of Classical Physics: What It Takes

With his usual wit John S. Bell once defined Dirac as "the most distinguished of 'why bother?'ers" (Bell 2004[2], p. 213). Although Bell referred to a popular article published by Dirac on *Scientific American* in the Sixties (Dirac 1963), even back at the time of his *Principles of Quantum Mechanics*, Dirac was far from taking the conceptual foundations of quantum mechanics as an issue worth pursuing. In this respect, the adequacy of quantum mechanics was not threatened in Dirac's view by questions like the status of the measurement process, so that the oscillations we have pointed out in the previous section could hardly become relevant. Heisenberg, on the contrary, had a different attitude toward the foundational questions, but in fact this turned out not to be sufficient to let the issue of the universality of quantum mechanics to occupy the center of the stage. Still in his Nobel Lecture (December 11th, 1933) Heisenberg writes:

> Bohr has shown in a series of examples how the perturbation necessarily associated with each observation indeed ensures that one cannot go below the limit set by the uncertainty relations. He contends that in the final analysis an uncertainty introduced by the concept of measurement itself is responsible for part of that perturbation remaining fundamentally unknown. (Heisenberg 1933, p. 298)

In a passage he appears to be well aware of the universality issue for quantum mechanics, but the very possibility of investigating seriously the implication at least of the universality as a working hypothesis is immediately closed:

> Since in connection with this situation *it is tempting to consider the possibility of eliminating all uncertainties by amalgamating the object, the measuring apparatuses, and the observer into one quantum-mechanical system*, it is important to emphasize that the act of measurement is necessarily visualizable, since, of course, physics is ultimately only concerned with the systematic description of space-time processes. The behaviour of the observer as well as his measuring apparatus must therefore be discussed according to the laws of classical physics, as otherwise there is no further physical problem whatsoever. Within the measuring apparatus, as emphasized by Bohr, all events in the sense of the classical theory will therefore be regarded as determined, this also being a necessary condition before one can, from a result of measurements, unequivocally conclude what has happened. In quantum theory, too, the scheme of classical physics which objectifies the results of observation by assuming in space and time processes obeying laws is thus carried through up to the point where the fundamental limits are imposed by the unvisualizable

---

of the opening sentence of the Heisenberg 1927 uncertainty paper we mentioned above, which treats *intuition* in a way that is similar to the way in which Dirac treats *picture*.

character of the atomic events symbolized by Planck's constant. (Heisenberg 1933, p. 298, emphasis added). [10]

Passages like this show that a major role has been played by the assumption of the indispensability of classical concepts and language, in order for the completion of a quantum measurement process to make sense, and we have reviewed how crucial in this direction the Bohr's influence on Heisenberg has been in the transition from the 1927 uncertainty paper to the 1930 book. So, putting aside the ambiguities and the oscillations induced by conceptions like the disturbance view, how should we interpret this notion of 'indispensability' in the overall Bohrian conceptual framework?

As recalled in the Introduction, the already convoluted view of Bohr concerning measurement in quantum mechanics has been long associated with a more or less tacit assumption, according to which the recourse to classical concepts and language, in order to account for the occurrence of definite outcomes after the measurement, is grounded on a truly ontological 'cut' between the classical and the quantum realms. For instance, John S. Bell voiced this attitude when he wrote:

> The founding fathers were unable to form a clear picture of things on the remote atomic scale. They became very aware of the intervening apparatus, and of the need for a 'classical' base from which to intervene on the quantum system. And so the shifty split.» (J.S. Bell, *Against measurement*, in *Speakable and Unspeakable in Quantum Mechanics*, CUP 2004², p. 228)

Still in 1998 we find a strong emphasis on such a classical/quantum cut in a work by a notable scholar like Peter Mittelstaedt. In his book *The Interpretation of Quantum Mechanics and the Measurement Process*, referring to the methodological assumptions of the Copenhagen interpretation, he attributes to Bohr "the hypothesis of the classicality of measuring instruments" an hypothesis whose character and implications are described as follows:

> This means that the apparatuses that are used for testing quantum mechanics must not only truly exist in the sense of physics, *but these apparatuses must also be macroscopic instruments that are subject to the laws of classical physics*. Consequently, the experimental outcomes of measurements are events in the sense of classical physics and can be treated by means of classical theories like mechanics, electrodynamics, etc. In this way, the strange and paradoxical features of quantum mechanics disappear completely in the measurement results, which can be thus described by means of classical physics and ordinary language. (Mittelstaedt 1998, p. 2, emphasis added).

---

[10] This is how Ernst Cassirer expresses the Heisenberg view in his 1936 book on determinism and indeterminism in modern physics: "[...] in the description of atomic phenomena a line has to be drawn between the observer's measuring apparatus and the object of observation. On both sides of this line, on the one which leads to the observer as well as on that which contains the object of observation, all processes are sharply determined: on this side by the laws of classical physics by which the measuring apparatus is to be described, on the other by the differential equations of quantum mechanics. But at the line itself there appears an uncertainty, since the influence of the means of observation on the object to be observed must be considered as a not completely controllable disturbance." (Cassirer 1956, p. 128).

In more recent times, this reading is developed by Henrik Zinkernagel. For intance, in his 2015 paper he refers to a passage in which Bohr argues that

> [...] in each case some ultimate measuring instruments, like the scales and clocks which determine the frame of space-time coordination – on which, in the last resort, even the definitions of momentum and energy quantities rest – must always be described *entirely* on classical lines, and consequently kept outside the system subject to quantum mechanical treatment." (Bohr 1938, p. 104, emphasis in the original).

According to Zinkernagel, one can make sense of this argument only under the assumption that quantum mechanics actually *fails* to be universal:

> A way to understand Bohr's requirement is that we need a reference frame to make sense of, say, the position of an electron (in order to establish with respect to what an electron has a position). And, by definition, a reference frame has a well-defined position and state of motion (momentum). Thus the reference frame is not subject to any Heisenberg uncertainty, and it is in this sense (and in this context) classical. This does not exclude that any given reference system could itself be treated quantum mechanically, but we would then need another – classically described – reference system e.g. to ascribe position (or uncertainty in position) to the former. (Zinkernagel 2015, p. 430).

A wide consensus has been growing in the years, however, according to which this view can hardly be maintainted[11]. As a matter of fact, even one of the textbooks that has been often considered among the most 'orthodox' in the Bohrian tradition – *Quantum Mechanics. Non-Relativistic Theory* by Lev Landau and Evgenij Lifšits, an influent treatise on non-relativistic quantum mechanics, whose first edition goes back to 1947 – suggested a more nuanced reading of the role of classical physics in the Bohrian analysis of the quantum measurement process. After emphasizing that "the importance of the concept of measurement in quantum mechanics was elucidated by N. Bohr", Landau and Lifšits write:

> It is in principle impossible, however, to formulate the basic concepts of quantum mechanics without using classical mechanics. [...] In this connection the "classical object" is usually called the *apparatus* and its interaction with the electron is spoken of as *measurement*. However it must be emphasized that we are here not discussing a process of measurement in which the physicist-observer takes part. By *measurement*, in quantum mechanics, we understand any process of interaction between classical and quantum objects occurring apart from and independently of any observer. [...] Thus quantum mechanics occupies a very unusual place among physical theories: it contains classical mechanics as a limiting case, yet at the same time it requires this limiting case for its own formulation. (Landau, Lifshitz 1956, pp. 2-3).

---

[11] A recent interpretation of quantum mechanics that reads Bohr in a non-universalistic fashion, in order to support its own account of the measurement process, is the most up-to-date version of the Bub information-theoretic interpretation (Bub 2018).

Not only the apparatus is indicated as "classical" in quotation marks (suggesting that this indication is just a *façon de parler*); equally important is the reference to the status of classical mechanics as a "limiting case for its own formulation", namely, as a tool whose essential role is simply that of making the very *formulation* of quantum mechanics possible.

That the recourse to classical physics is unavoidable in order for quantum mechanics to be about anything at all, in Bohr's view, is already apparent from a passage taken from the very 1928 complementarity paper:

> According to the above considerations regarding the possibilities of definition of the properties of individuals, it will obviously make no difference in the discussion of the accuracy of the measurements of position and momentum of a particle if collisions with other material particles are considered instead of scattering of radiation. *In both case, we see that the uncertainty in question equally affects the description of the agency of observation and the object. In fact, this uncertainty cannot be avoided in a description of the behaviour of individuals with respect to coordinate system fixed in the ordinary way by means of solid bodies and unperturbable clocks*. (Bohr 1928, p. 66, emphasis added).

This passage suggests a sort of universality thesis for quantum mechanics ("we see that the uncertainty in question *equally affects* the description of the agency of observation and the object") and, at the same time, paves the way for the necessity of classical language and concepts in order for the measurement process to make sense ("this uncertainty cannot be avoided in a description of the behaviour of individuals with respect to coordinate system fixed in the ordinary way by means of solid bodies and unperturbable clocks."). The latter point led for instance such a distinguished Bohr scholar as Dugald Murdoch to consider the measurement interaction as 'indeterminable' (Murdoch 1987, p. 97) in purely quantum-mechanical terms, under the general assumption of Bohr's philosophy of physics:

> Bohr may also believe that treatment of the instrument in classical terms is necessary not only if observation is to be objective, or even possible at all, but also if it is to yield a *definite*, unambiguous result. If the instrument is treated in quantum-mechanical terms, then after the interaction it is generally described by a state vector which is a superposition of states pertaining to the measured observable. The result of the measurement is definite only if the instrument is treated in classical physical terms [...]. This is what Bohr had in mind when he says that 'only with the help of classical ideas it is possible to ascribe an unambiguous meaning to the results of observations'. (Murdoch 1987, p. 98) [12]

As I anticipated in the summary, two are the main points of the Bohr analysis of the quantum measurement process that I would like to recall, in view of the discussion of the next section.

---

[12] In a defence of the view, according to which the Bohr texts would not justify an interpretation of his thought to the effect that there exists an independent natural realm of an intrinsic classical character, Landsman had argued for instance: "there is no doubt that both Bohr and Heisenberg believed in the fundamental and universal nature of quantum mechanics, and saw the classical description of the apparatus as a purely *epistemological* move, which expressed the fact that a given quantum system is being used as a measuring device" (Landsman 2007, p. 437, emphasis added).

First of all, why are we justified in claiming that Bohr has an outstanding role in specifically and thematically addressing the relation classical/quantum in the quantum measurement process as a foundational issue in its own right? Second, on what basis can we see the Bohrian conceptual framework as supporting a *coexistence* of the use of the language of classical physics as a pragmatic tool on one side and the universality of quantum mechanics as a theory governing the whole physical world on the other?

In fact, the two questions are to be treated unitarily, since the need to justify the coexistence referred to in the second point emerges only as a consequence of the development of a robust (and not only pragmatic) view on the first point. To begin with, there is no better strategy than using again the complementarity paper, whose section 1 can well be interpreted in the direction of a universality thesis for quantum mechanics:

> Indeed, our usual description of physical phenomena is based entirely on the idea that the phenomena concerned may be observed without disturbing them appreciably. This appears, for example, clearly in the theory of relativity, which has been so fruitful for the elucidation of the classical theories. As emphasised by Einstein, every observation or measurement ultimately rests on the coincidence of two independent events at the same space-time point. Just these coincidences will not be affected by any differences which the space-time co-ordination of different observers otherwise may exhibit. *Now the quantum postulate implies that any observation of atomic phenomena will involve an interaction with the agency of observation not to be neglected. Accordingly, an independent reality in the ordinary physical sense can neither be ascribed to the phenomena nor to the agencies of observation*. (Bohr 1934, pp. 53-54)

Several further passages can be quoted for our purposes, as many interpreters have noted. Take for instance an article, "Causality and Complementarity", derived from an address delivered in 1936 at the Second International Congress for the Unity of Science in Copenhagen. Here Bohr argues that what he calls "the assumption underlying the ideal of causality", namely "that the behavior of a physical object relative to a given system of coordinates is uniquely determined, quite independently of whether it is observed or not" has been preserved by the theory of relativity but *not* by quantum mechanics. In addressing the question of the role of observation in quantum terms, Bohr writes:

> [...] a still further revision of the problem of observation has since been made necessary by the discovery of the universal quantum of action, which has taught us that the whole mode of description of classical physics, including the theory of relativity, retains its adequacy only as long as all quantities of action entering into the description are large compared to Planck's quantum. When this is not the case, as in the region of atomic physics, there appear new uniformities which cannot be fitted into the frame of the ordinary causal description. *This circumstance, at first sight paradoxical, finds its elucidation in the recognition that in this region it is no longer possible sharply to distinguish between the autonomous behavior of a physical object and its inevitable interaction with other bodies serving as measuring instruments, the direct consideration of which is excluded by the very nature of the concept of observation itself*. (Bohr 1937, p. 290, emphasis added)

In some sense, therefore, it is the very nature of observation in the microphysical realm that forces us to treat "the autonomous behavior of a physical object and its inevitable interaction with other bodies serving as measuring instruments" as a unitary process wholly governed by the laws of quantum mechanics. On the other hand, the measuring instruments play what has been called an 'epistemic function' (Camilleri 2017), namely that of allowing to extract empirical information from phenomena, and it is in this respect that the issue of the indispensability of classical physics' concepts arises. As we will read in the 1938 Bohr paper that we will treat in more detail in the next section, in this epistemic function "the quantum mechanical treatment [of the measurement agency] will for this purpose be *essentially equivalent* with a classical description" (Bohr 1938, p. 104, emphasis added): this expression clearly suggests that, *in principle*, both the measured system and the measurement agency are indeed quantum systems but also that, *in practice* and for the sake of the epistemic function that the measurement process is supposed to fulfil, the outcome of experiments will have to receive a classical account.

## 4  Bohr and von Neumann: from the 1932 von Neumann Model of the Quantum Measurement to the 1938 article "The causality problem in atomic physics"

On the basis of a wide consensus, whose main reasons have been summarized in the preceding section, we may take for established that Bohr *did* believe in the fundamental and universal nature of quantum mechanics and that the use of classical concepts and language in describing the "agencies of observations" concerns just epistemology and not ontology. Since the von Neumann formal model of measurement, as developed in his 1932 treatise, is an explicit 'implementation' of the thesis on the universal nature of quantum mechanics, a question is bound to emerge quite naturally: what is the status of the Bohr theory of quantum measurement *with respect to this model*? More generally, what can we learn from a more thorough investigation on the relations between the Bohr and von Neumann respective models of measurement? This point is in fact still virtually ignored but, in my opinion, should deserve the utmost attention. Bohr can well be credited with the awareness that, when we move from the mere pragmatics of experiment to the foundations of the theory, the system/apparatus correlation is to be addressed at the fundamental level in a *quantum* context; as a consequence, the contemporary scholarship on the Bohr philosophy of physics might find worth addressing the question: how does the Bohrian view of the measurement process in quantum mechanics fare on the background of the von Neumann rigorous treatment of measurement in 'universalistic' terms? The arguments that I will propose in the present section are meant to contribute in this direction.

In the entire, monumental Bohr biography by Abraham Pais, all we find about the Bohr-von Neumann relationship is just one single reference to von Neumann:

> I do know, though, that when in Princeton Bohr would often discuss measurement theory with Johnny von Neumann who pioneered that field. A little of those discussions has even appeared in print." (Pais 1991, p. 435).

Pais is right in suggesting that hardly anything of the discussions between Bohr and von Neumann even appeared in print, since *de facto* the only occasion of this kind seems to have been the Bohr contribution to a conference, held in Warsaw between the 30th of May and the 3rd of June 1938 and entitled *New Theories in Physics* (Bohr 1938). In a footnote Pais mentions this conference as the circumstance in which Bohr and von Neumann found themselves together in a public scientific event. In addition to them, other leading figures in contemporary physics attended the conference, such as Arthur Eddington, Léon Brillouin, Louis De Broglie, George Gamow, Paul Langevin, Léon Rosenfeld, Samuel Goudsmit, Eugene Wigner. The analysis of the Bohr 1938 paper shows that he elaborates specifically on the von Neumann model for his own purposes, hence this work is crucial in order to shed some light on the Bohr-von Neumann relationship concerning the analysis of the quantum measurement process: to see why, let me first review the main elements of the von Neumann model of a quantum measurement.

The first analysis of the quantum measurement process in a direction that could pave the way to the hypothesis of the universality of quantum mechanics is probably a paper by von Neumann on the statistical structure of the theory (von Neumann 1928). In a review paper published in 1958 in the *Bulletin of the. American Mathematical Society*, the Belgian physicist Léon van Hove stressed the importance of that pioneering paper, but mentioned a connection with Niels Bohr that in fact is quite obscure in van Hove terms:

> In the same paper von Neumann also investigates a problem which is still now the subject of much discussion, viz., the theoretical description of the quantum-mechanical measuring process and of the noncausal elements which it involves. Mathematically speaking von Neumann's study of this delicate question is quite elegant. *It provides a clear-cut formal framework for the numerous investigations which were needed to clarify physically the all-important implications of quantum phenomena for the nature of physical measurements, the most essential of which is Niels Bohr's concept of complementarity*. (van Hove 1958, p. 97, emphasis added).

The 1928 results were included by von Neumann in a wider framework in his 1932 treatise on the mathematical foundations of quantum mechanics[13], in which the assumption of existence within the theory of two distinct modalities of evolution over time for the states of quantum systems was formalized. The first is the unitary evolution (called "causal change", for instance, at page 417) governed by the time-dependent deterministic Schrödinger equation – holding between measurements – whereas the second is the non-unitary, irreversible, stochastic evolution induced by measurements. In Chapter III ("The quantum statistics") von Neumann describes the latter modality of evolution as follows. Let $R$ a self-adjoint operator representing a physical quantity ℝ, to be measured on a quantum system in the state $\psi$. Under standard mathematical constraints on $R$,

> a measurement of ℝ has the consequence of changing each state $\psi$ into one of the states $\phi_1$, $\phi_2$,..., which are connected with the respective results of measurements $\lambda_1$, $\lambda_2$,... The probabilities of these changes are therefore equal to the measurement probabilities for $\lambda_1$, $\lambda_2$,... (von Neumann 1955, p. 216).

---

[13] All quotes and page references are taken from the English edition (von Neumann 1955).

Few lines later, von Neumann expresses clearly his uneasiness in characterizing the nature of this transition:

> We have then answered the question as to what happens in the measurement of a quantity ℝ, under the above assumptions for its operator $R$. *To be sure, the "how" remains unexplained for the present.* This discontinuous transition from $\psi$ into one of the states $\phi_1$, $\phi_2$,..., [...] is certainly not of the type described by the time dependent Schrödinger equation. (von Neumann 1955, p. 217, emphasis added).

It is important to remark that when, in his 1930 book, Dirac introduces what has become known as the 'collapse postulate', he does not feel compelled to mention any lack of explanation as to the status of this 'collapse'. After describing what happens in a quantum measurement when we measure a second time a physical quantity that has been measured in an immediately preceding time, Dirac simply states: "In this way we see that a measurement always *causes* the system to jump into an eigenstate of the dynamical variable that is being measured, the eigenvalue this eigenstate belongs to being equal to the result of the measurement." (Dirac 1958[4], p. 36). On the other hand von Neumann promises to address the problem of the "how": "we shall attempt to bridge this chasm later" (von Neumann 1955, p. 217).

The whole Chapter VI, the last of the book, is in fact entirely devoted to this attempt, whose formulation is in some sense 'prepared' by some remarks in the section 1 of the Chapter V. Recalling once again the coexistence of "arbitrary changes" and "automatic changes" (von Neumann 1955, p. 351), i.e. of the two modalities of evolution in quantum mechanics – (**1.**) denotes the non-unitary evolution, whereas (**2.**) denotes the unitary, Schrödinger evolution – von Neumann writes:

> First of all, it is noteworthy that the time dependence of $H$ is included in **2.**, so that one should expect that **2.** would suffice to describe the intervention caused by a measurement: indeed, a physical intervention can be nothing else than the temporary insertion of a certain energy coupling into the observed system, i.e. the introduction of an appropriate time dependency of *H* (prescribed by the observer). Why then do we need the special process **1**. for the measurement? The reason is this: in the measurement we cannot observe the system **S** by itself, but must rather investigate the system **S** + **M**, in order to obtain (numerically) its interaction with the measuring apparatus **M**. *The theory of the measurement is a statement concerning* **S** + **M**, *and should describe how the state of **S** is related to certain properties of the state of **M*** (namely, the positions of a certain pointer, since the observer reads these). (von Neumann 1955, p. 352, emphasis added)

The emphasized lines apparently express the endorsement of a universality thesis for quantum mechanics, namely they make it clear that (i) the account of the measurement process is an account of the system **S** + **M**, and (ii) this account is to be given *entirely* in quantum mechanical terms. Not just that:

> Moreover, *it is rather arbitrary* whether or not one includes the observer in **M**, and replaces the relation between the **S** state and the pointer positions in **M** by the relations of this state and the chemical changes in the observer's eye or even in the brain (i.e. , to that which he has "seen" or "perceived"). We shall investigate this more precisely in VI.1. In any case, therefore, the application of **2**. is of importance only for **S** + **M**. *Of course, we must show that this gives the same result for* **S** *as the direct application of* **1**. *on* **S**. *If this is successful, then we have achieved a unified way of looking at the physical world on a quantum mechanical basis*. (von Neumann 1955, p. 352, emphasis added).

Passages like this might be echoing what Bohr had already stated along similar lines in his 1928 complementarity article:

> In tracing observations back to our sensations, once more regard has to be taken to the quantum postulate in connection with the perception of agency of observation, be it through its direct action upon the eye or by means of suitable auxiliaries such as photographic plates, Wilson clouds, etc. It is easily seen, however, that the resulting additional statistical element will not influence the uncertainty in the description of the object." (Bohr 1928, p. 67).

Be it as it may, the bulk of the von Neumann treatment of the quantum measurement process in the Chapter VI is exactly to prove that this conjecture holds true, namely that what has been called the 'consistency problem' (Bub 2001, p. 65) for the two modalities of evolution can be solved.

In introducing the problem, von Neumann argues for the necessity to address the problem of the role of *subjectivity*:

> It is inherently correct that the measurement or the related process of the subjective perception is a new entity relative to the physical environment and is not reducible to the latter. Indeed, subjective perception leads us into the intellectual inner life of the individual, which is extra-observational by its very nature (since it must be taken for granted by any conceivable observation or experiment). (von Neumann 1955, p. 418).

This 'extra-observational' subjectivity notwithstanding, it is necessary to provide a correlational formulation of subjective perceptions, so to say, namely to express somehow the content of these perceptions in terms of correlations between outcomes and recording procedures. This is the assumption von Neumann dubs as *principle of psycho-physical parallelism*, something he takes to be "a fundamental requirement of the scientific viewpoint" (von Neumann 1955, p. 418):

> it must be possible so to describe the extra-physical process of the subjective perception as if it were in reality in the physical world – i.e., to assign to its very parts equivalent processes in the objective environment, in ordinary space.

So, how does von Neumann formulate the consistency problem so as to respect, at the same time, the principle of psycho-physical parallelism? Let us divide the world in three parts:

1. the observed system $\mathbf{S}$,
2. the measuring apparatus $\mathbf{M}$,
3. the observer $\mathbf{O}$.

Moreover, under the assumption of universality for quantum mechanics, let us suppose two different ways of studying the interaction between $\mathbf{S}$, $\mathbf{M}$ and $\mathbf{O}$, namely, either via the composition

$$\mathbf{S} + [\mathbf{M} + \mathbf{O}]$$

or via the composition

$$[\mathbf{S} + \mathbf{M}] + \mathbf{O}$$

so that the consistency problem is (von Neumann 1955, p. 420): does the application of process **1.** (the non-unitary evolution) to $\mathbf{S}$ (as measured by $\mathbf{M} + \mathbf{O}$) or to $\mathbf{S} + \mathbf{M}$ (as measured by $\mathbf{O}$) change the statistics *for $\mathbf{S}$* ?

The answer turns out to be *negative*: as succinctly put by Bub,

> applying process 1 directly to $[\mathbf{S}]$ yields the same density operator for the statistics of all $[\mathbf{S}]$-observables after the measurement as we obtain by considering the measurement as a process 2 interaction between $[\mathbf{S}]$ and $[\mathbf{M}]$, and then applying process 1 to the measurement of $[\mathbf{S} + \mathbf{M}]$ by a second instrument $[\mathbf{O}]$. (Bub 2001, p. 67).

But what is then the conceptual implication of the solution of the consistency problem? If this problem had turned out to be unsolvable, fixing the boundary between $\mathbf{S}$ and $[\mathbf{M} + \mathbf{O}]$ in the composition $\mathbf{S} + [\mathbf{M} + \mathbf{O}]$ or between $[\mathbf{S} + \mathbf{M}]$ and $\mathbf{O}$ in the composition $[\mathbf{S} + \mathbf{M}] + \mathbf{O}$ would not be conventional anymore; as a consequence, this would force us to find physically serious reasons to discover *where exactly* should we draw the boundary and, in turn, this would affect the psycho-physical parallelism, hence a serious blow to the general scientific viewpoint.

Therefore, according to von Neumann, the actual existence of a solution is highly relevant because it turns out to be a mathematical proof that the act of drawing the boundary either between **S** and [**M** + **O**] or between [**S** + **M**] + **O** is a conventional act, devoid of any physical meaning and as such of a totally pragmatic character: it is exactly this character that in von Neumann's view allows us to claim that "we have achieved a unified way of looking at the physical world on a quantum mechanical basis."

If this reconstruction is correct, two important points follow. First, the von Neumann argument on the virtue of the above mentioned arbitrariness is in tension with the claim that von Neumann had a firm belief in the collapse of the wave function as a 'real' physical process: should we take seriously the idea that there is something *real* that *physically* collapses in the measurement interaction, how can the fixation of the boundary either between **S** and [**M** + **O**] or between [**S** + **M**] + **O** be really arbitrary?[14] Second, the virtue that von Neumann ascribes to the above mentioned conventionality might have been taken by Bohr to be a mathematically robust confirmation of his position on the *pragmatic* justification of using classical concepts for treating the quantum measurement process. That is why I argue that the von Neumann claim – according to which solving the consistency problem is equivalent to achieve a unified way of looking at the physical world 'on a quantum mechanical basis' – played the role of a motivating factor for the Bohrian account of measurement in the Thirties: in order to see in which terms this might have been the case, let me turn now to the paper that Bohr delivered at the 1938 conference – entitled "The causality problem in atomic physics".

Bohr devotes the section 2 to what he calls "The observation problem in quantum theory" (Bohr 1938, p. 99): a textual analysis of this section shows (I argue) a close correspondence with the von Neumann framework, as displayed in the above mentioned Chapter VI of *Die Mathematische Grundlagen*. First of all, let us consider the very definition of the notion of measurement provided by Bohr:

> In the first place we must recognize that a measurement can mean nothing else than the unambiguous comparison of some property of the object under investigation with a corresponding property of another system, serving as a measurement instrument, *and for which this property is directly determinable according to its definition in everyday language or in the terminology of classical physics*. (Bohr 1938, p. 100).

I take this passage to be a sort of re-formulation of the von Neumann psycho-physical parallelism where, *à la* Bohr, the role of the 'subjective perception' is taken over by the 'everyday language', in conformity to the Bohrian attitude of purging the analysis of measurement from any subjectivistic and 'mentalistic' emphasis. Once room has been made to the *pragmatic* necessity of adopting the everyday language and the terminology of classical physics, Bohr proceeds in a direction which is substantially coherent with a universalistic attitude:

> While within the scope of classical physics such a comparison can be obtained without intefrering essentially with the behaviour of the object, this is not so in the field of quantum theory, where the interaction between the object and the measuring instruments will have an essential influence *in the phenomenon itself*. (Bohr 1938, p. 100).

---

[14] For a sustained argument against the view, according to which von Neumann holds that the measurement process produces a physical collapse, see Becker 2004.

Remarkably, Bohr refers here to what is to be described by quantum theory as a 'phenomenon', namely as something that the conditions of the measurement-as-interaction contribute to *constitute*[15]. It is essentially on the basis of this 'constitutive' role of the measurement interaction that we may explain the Bohr's hostility toward the disturbance view of measurement that we discussed above:

> The unaccostomed features of the situation with which we are confronted in quantum theory necessitate the greatest caution as regards all questions of terminology. *Speaking, as is often done, of disturbing a phenomenon by observation, or even of creating physical attributes to objects by measuring processes, is, in fact, liable to be confusing, since all such sentences imply a departure from basic conventions of language which, even though it sometimes may be practical for the sake of brevity, can never be ambiguous*. It is certainly far more in accordance with the structure and interpretation of the quantum mechanical symbolism, as well as with elementary epistemological principles, to reserve the word "phenomenon" for the comprehension of the effects observed under given experimental conditions. (Bohr 1938, p. 104, emphasis added)

Moreover, such a constitutive attitude is decisively more compatible with a universalistic view, and justifies the relation of non-separability between measured system and measuring instrument (precurring the quantum 'contextuality' discussed decades later) that Bohr takes to be at the heart of the quantum theory of measurement:

> The essential lesson of the analysis of measurements in quantum theory is thus the emphasis on the necessity, in the account of the phenomena, of taking the whole experimental arrangement into consideration, *in complete conformity with the fact that all unambiguous interpretation of the quantum mechanical formalism involves the fixation of the external conditions, defining the intial state of the atomic system concerned and the character of the possible predictions as regards subsequent observable properties of that system*. (Bohr 1938, p. 101, emphasis added).

My conjecture here is that the emphasized lines – with the non accidental reference to the 'quantum mechanical formalism' – represent a (typically idiosyncratic) Bohrian wording of the von Neumann description of a measurement of a physical quantity on a system **S** by an apparatus **M** in terms of the entirely quantum-mechanical joint system **S** + **M**. Moreover, I claim it can be safely shown that Bohr, few pages later, also acknowledges the von Neumann solution of the consistency problem outlined above, a solution according to which the statistics for the measured system **S** is unaffected by possible different choices of couplings with the measuring apparatus **M** and the observer **O**:

> In the system to which the quantum mechanical formalism is applied, it is of course possible to include any intermediate auxiliary agency employed in the measuring process. Since, however, all those properties of such agencies which, according to the aim of the measurement, have to be compared with corresponding properties of the object, must be described on classical lines, *their quantum mechanical treatment will for this purpose be essentially equivalent with a classical description*. The question of eventually including such agencies within the system under investigation is thus *purely a matter of practical convenience*, just as in classical physical measurements; and such displacements of the section between object and measuring instruments can therefore never involve any arbitrariness in the description of a phenomenon and its quantum mechanical treatment. The only significant point is that in each case some ultimate measuring instruments [...] must always be described entirely on classical lines, and consequently kept outside the system subject to quantum mechanical treatment. (Bohr 1938, p. 104, emphasis added). [16]

---

[15] For arguments supporting such constitutive view in the Bohrian conceptual framework, see for instance Kaiser 1992, Cuffaro 2010, Bitbol 2017.

So, on the one hand, the addition of 'intermediate auxiliary agencies' creates no problem to the theory, which is capable to handle in a mathematically sound way the correlation among them and no confusion may derive by the possibility of displacing at different points the (conventional) line between system and observer. The Bohr reference to this possible addition resonates quite precisely with the von Neumann remark, quoted above, according to which "it is rather arbitrary whether or not one includes the observer in **M**, and replaces the relation between the **S** state and the pointer positions in **M** by the relations of this state and the chemical changes in the observer's eye or even in the brain (i.e. , to that which he has "seen" or "perceived")." (von Neumann 1955, p. 352). On the other hand, as a consequence of the choice that in each concrete case has to be made on a specific location of this line, there will some 'ultimate' instrument that, for pragmatic reasons of inter-subjectivity and communication of the experiment results, will be treated in terms of classical physics and its language, consistently with the pragmatic indispensability of this language assumed by Bohr. In the 1938, therefore, Bohr appears to be aware of the possibility of exploiting von Neumann's work for his foundational purposes. In particular, the conventionality that in the von Neumann treatment makes unproblematic the conventional fixation of the boundaries of the couplings among **S**, **M** and **O** is 'put to work' by Bohr in *his* framework, in order to make unproblematic the above use of the language of classical physics in the analysis of quantum measurement, also on the basis of the von Neumann mathematical authority: another bit of evidence that, as Pais put it succintly, indeed Bohr *did* "discuss measurement theory with Johnny von Neumann who pioneered that field".

---

[16] Thus it appears far from accidental that, in the section 1, Bohr had already emphasized that "the completeness and *self-consistency* of the whole formalism is most clearly exhibited by the elegant axiomatic exposition of von Neumann" (Bohr 1938, p. 98).

# 5    Conclusions

It is nowadays widely recognized that the acclaimed status of father of the Copenhagen interpretation did to Niels Bohr more harm than good. Abner Shimony once confessed that "after 25 years of attentive – and even reverent – reading of Bohr, I have not found a consistent and comprehensive framework for the interpretation of quantum mechanics" (Shimony 1985, p. 109) and an unsympathetic scholar such as Mara Beller wrote that "even the most competent and friendly readers find Bohr's philosophy obscure and inconsistent" (Beller 1999, p. 275). Due to a wealth of historical and analytic work that has been done in more recent years, however, outlining a more balanced and sophisticated view of the Bohr thought on quantum mechanics seems now an aim in sight. [17] In the present paper I have attempted to do some justice to the Bohr view of the quantum measurement process, in particular by investigating its connections with the von Neumann work in the mathematical foundations of quantum mechanics. In this respect, I have tried to study in Bohr's thought the coexistence of the assumption of the classical physics and language indispensability with a universality thesis for quantum mechanics, in light of what I take as a plausible hypothesis: since von Neumann conceived *explicitly* and *from the start* the analysis of the measurement process in microphysics as the analysis of a process *entirely* governed by quantum laws – a view according to which, at a fundamental level, all physical phenomena are *quantum* phenomena – Bohr *did* try to make this analysis and his own compatible. In this respect, the core of the present paper is to show in what sense Bohr seriously attempted to locate his view of the quantum measurement process within the mathematically robust formulation that von Neumann did so much to consolidate. If successful, my analysis would contribute to substantiate a picture in which Bohr, far from representing the quantum novelty in terms of an unclear notion of 'disturbance' that classical physics was unable to encompass (a notion still present in major representatives of the quantum mechanics' community such as Heisenberg and Dirac), takes seriously the implications of the *whole* empirical world being quantum in nature.

---

[17] For instance, in a contribution on the alleged 'obscurity' of Bohr, Camilleri argued interestingly that we should see the Bohrian conceptual framework not as an attempt to provide an interpretation *of quantum mechanics*, but rather as a *philosophy of the quantum experiment* (Camilleri 2017).